%% file: ms.tex
\documentclass[journal=ancac3,manuscript=article,layout=twocolumn]{achemso}
\pdfoutput=1
\newcommand*{\citen}[1]{%
  \begingroup
    \romannumeral-`\x 
    \setcitestyle{numbers}%
    \cite{#1}%
  \endgroup   
}
\usepackage[cm]{sfmath}
\usepackage{float}
\usepackage{blindtext} 
\usepackage[version=3]{mhchem} 
\usepackage[utf8]{inputenc}
\usepackage{amsmath}

\author{Meike Reginka}
\affiliation[AGE]{Institute of Physics and Center for Interdisciplinary Nanostructure Science and Technology (CINSaT), University of Kassel, Heinrich-Plett-Strasse 40, D-34132 Kassel (Germany)}
\alsoaffiliation[AIMED]{Artificial Intelligence Methods for Experiment Design (AIM-ED), Joint Lab Helmholtzzentrum für Materialien und Energie, Berlin (HZB) and Kassel University, cc Gregor Hartmann, Hahn-Meitner Platz 1, 14109 Berlin, Germany}
\email{reginka@uni-kassel.de}
\author{Hai Hoang}
\affiliation[AGE]{Institute of Physics and Center for Interdisciplinary Nanostructure Science and Technology (CINSaT), University of Kassel, Heinrich-Plett-Strasse 40, D-34132 Kassel (Germany)}
\author{Özge Efendi}
\affiliation[BC]{Institute of Biology and Center for Interdisciplinary Nanostructure Science and Technology (CINSaT), University of Kassel, Heinrich-Plett-Strasse 40, D-34132 Kassel (Germany)}
\author{Maximilian Merkel}
\affiliation[AGE]{Institute of Physics and Center for Interdisciplinary Nanostructure Science and Technology (CINSaT), University of Kassel, Heinrich-Plett-Strasse 40, D-34132 Kassel (Germany)}
\alsoaffiliation[AIMED]{Artificial Intelligence Methods for Experiment Design (AIM-ED), Joint Lab Helmholtzzentrum für Materialien und Energie, Berlin (HZB) and Kassel University, cc Gregor Hartmann, Hahn-Meitner Platz 1, 14109 Berlin, Germany}
\author{Rico Huhnstock}
\affiliation[AGE]{Institute of Physics and Center for Interdisciplinary Nanostructure Science and Technology (CINSaT), University of Kassel, Heinrich-Plett-Strasse 40, D-34132 Kassel (Germany)}
\alsoaffiliation[AIMED]{Artificial Intelligence Methods for Experiment Design (AIM-ED), Joint Lab Helmholtzzentrum für Materialien und Energie, Berlin (HZB) and Kassel University, cc Gregor Hartmann, Hahn-Meitner Platz 1, 14109 Berlin, Germany}
\author{Dennis Holzinger}
\affiliation[AGE]{Institute of Physics and Center for Interdisciplinary Nanostructure Science and Technology (CINSaT), University of Kassel, Heinrich-Plett-Strasse 40, D-34132 Kassel (Germany)}
\author{Kristina Dingel}
\affiliation[IT]{Intelligent Embedded Systems, University of Kassel, Wilhelmshöher Allee 71-73, D-34121 Kassel (Germany)}
\alsoaffiliation[AIMED]{Artificial Intelligence Methods for Experiment Design (AIM-ED), Joint Lab Helmholtzzentrum für Materialien und Energie, Berlin (HZB) and Kassel University, cc Gregor Hartmann, Hahn-Meitner Platz 1, 14109 Berlin, Germany}
\author{Bernhard Sick}
\affiliation[IT]{Intelligent Embedded Systems, University of Kassel, Wilhelmshöher Allee 71-73, D-34121 Kassel (Germany)}
\alsoaffiliation[AIMED]{Artificial Intelligence Methods for Experiment Design (AIM-ED), Joint Lab Helmholtzzentrum für Materialien und Energie, Berlin (HZB) and Kassel University, cc Gregor Hartmann, Hahn-Meitner Platz 1, 14109 Berlin, Germany}
\author{Daniela Bertinetti}
\affiliation[BC]{Institute of Biology and Center for Interdisciplinary Nanostructure Science and Technology (CINSaT), University of Kassel, Heinrich-Plett-Strasse 40, D-34132 Kassel (Germany)}
\author{Friedrich W. Herberg}
\affiliation[BC]{Institute of Biology and Center for Interdisciplinary Nanostructure Science and Technology (CINSaT), University of Kassel, Heinrich-Plett-Strasse 40, D-34132 Kassel (Germany)}
\email{herberg@uni-kassel.de}
\author{Arno Ehresmann}
\affiliation[AGE]{Institute of Physics and Center for Interdisciplinary Nanostructure Science and Technology (CINSaT), University of Kassel, Heinrich-Plett-Strasse 40, D-34132 Kassel (Germany)}
\alsoaffiliation[AIMED]{Artificial Intelligence Methods for Experiment Design (AIM-ED), Joint Lab Helmholtzzentrum für Materialien und Energie, Berlin (HZB) and Kassel University, cc Gregor Hartmann, Hahn-Meitner Platz 1, 14109 Berlin, Germany}

\title[BFP Paper]
{Transport efficiency of biofunctionalized magnetic particles tailored by surfactant concentration}
\keywords{particle transport, magnetic bead, exchange bias, IBMP, magnetic field landscape, biofunctionalization, green fluorescent protein (GFP), Lab-on-chip (LOC), surface forces \newline}
\begin{document}

\begin{abstract}
Controlled transport of surface functionalized magnetic beads in a liquid medium is a central requirement for the handling of captured biomolecular targets in microfluidic lab-on-chip biosensors. Here, the influence of the physiological liquid medium on the transport characteristics of functionalized magnetic particles and on the functionality of the coupled protein is studied. These aspects are theoretically modeled and experimentally investigated for prototype superparamagnetic beads, surface functionalized with green fluorescent protein immersed in buffer solution with different concentrations of a surfactant. The model reports on the tunability of the steady-state particle substrate separation distance to prevent their surface sticking via the choice of surfactant concentration. Experimental and theoretical average velocities are discussed for a ratchet like particle motion induced by a dynamic external field superposed on a static locally varying magnetic field landscape. The developed model and experiment may serve as a basis for quantitative forecasts on the functionality of magnetic particle transport based lab-on-chip devices.
\end{abstract}


\input{Introduction}
\section{Model}
\input{Model}
\section{Results and Discussion}
\input{Results}
\section{Conclusions}
\input{Conclusion}
\section{Experimental Methods}
\input{Experimental}

\begin{acknowledgement}
The authors thank the Center for Interdisciplinary Nanostructure Science and Technology (CINSaT) at Kassel university promoting cross-disciplinary communication and research, project "`MASH"' supported by an internal grant of Kassel university. Moreover, we acknowledge the participation of J. Rühl in transport experiments.
\end{acknowledgement}
%
\begin{suppinfo}
Additional information for the model part, about the coupling efficiency of GFP and the evaluation of particle velocities is given.
\end{suppinfo}


\bibliography{bibliography}

\end{document}


\input{Supplementary}
\bibliography{bibliography}

%% file: Introduction.tex
Lab-on-a-chip\cite{Knight2002} tests are in the focus of medical diagnostics, where disease-relevant biomarkers may be detected in human body fluids (\textit{e.g.} blood serum, urine). For these systems, magnetic micro- or nanobeads (in the following synonymously referred to as particles) are discussed to play a key role for single to few molecule detection. In on-chip biosensors, functionalized particles may cover functionalities like stirring, analyte capturing, transport, and detection,\cite{Issadore2014,Lim2017a,Holzinger2012,Owen2016} where for the latter they may serve as labels for magnetic sensing technologies like magnetoresistive elements.\cite{Donolato2009,Lim2017,Lin2017,Weddemann2010,Rampini2016} Precise bead motion control therefore is the basis for 
these functionalities.\cite{Lim2017a, Holzinger2012, Owen2016} 
Corresponding actuation concepts for superparamagnetic beads (SPBs) are based on magnetic thin film systems which are topographically\cite{Yellen2007} or magnetically patterned\cite{Holzinger2015,Donolato2011,Tierno2008,Mirzaee-Kakhki2020} into designed micromagnetic elements for SPB guidance by magnetic stray field tracks.\cite{Ehresmann2015, Rampini2016} The tracks can, \textit{e.g.}, be achieved by moving domain walls\cite{Donolato2011,Urbaniak2010} or by the design of asymmetric magnetic potentials, commonly referred to as magnetic ratchets.\cite{Auge2009,Weddemann2010}

Hence, dynamically transforming the SPBs' magnetic potential energy landscape by the application of periodic external magnetic fields leads to their directed transport as the beads follow the shifted energy minima.\cite{Holzinger2015,Yellen2007} Although the local stray field strengths originating from these magnetic domain configurations are not very strong, they possess high field gradients which cause high transport velocities in external fields below 2~kA/m.\cite{Holzinger2015} For the design of the described magnetic stray field landscapes from the magnetic or topographic patterns towards specific properties of the SPB transport, micromagnetic simulations are frequently used to estimate the magnetic potential energy of the individual beads and, thus, allow for the prediction of their motion.\cite{Lim2017,Ouk2017,Holzinger2015,Yellen2007} 

Although it is obvious that the SPBs' motion in liquid environment is governed mainly by the force exerted by the magnetic potential energy landscape and consequently the drag force,\cite{Holzinger2015,Wirix-Speetjens2005} there are multiple hidden factors influencing the motion that have not been profoundly considered so far. Interestingly, the impact of surface forces between magnetic beads and substrate surface has only been discussed for non-functionalized SPBs in water,\cite{Wirix-Speetjens2005} however, they determine the colloidal stability and the sticking probability to the substrate surface of the SPBs and define, therefore, the necessary conditions for a SPB transport in the respective liquid. 
\begin{figure}[tb]%
\includegraphics[width=\columnwidth]{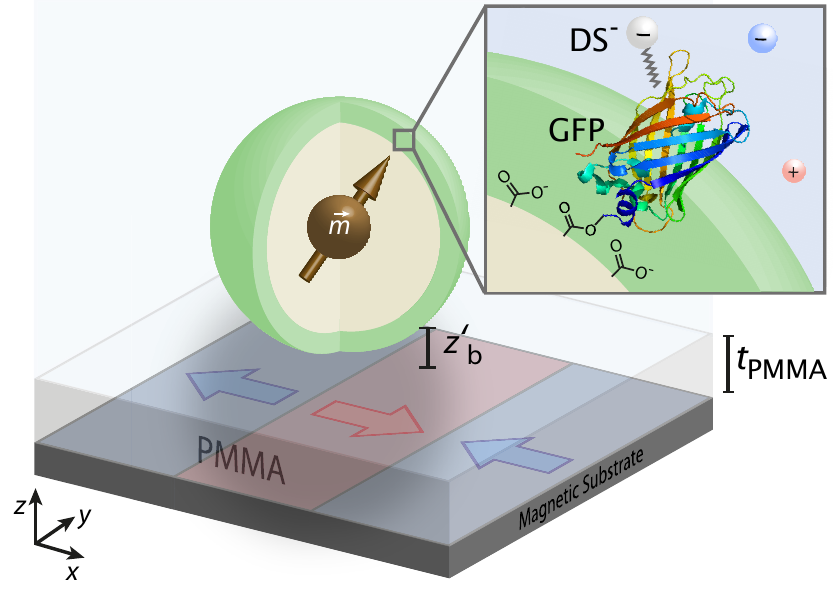}%
\caption{Sketch of the prototype system for the investigation of the colloidal stability and surface adhesion of superparamagnetic core-shell particles with biochemically immobilized green fluorescent protein (GFP), not drawn to scale. The substrate consists of a parallel stripe magnetic domain pattern with in-plane opposing magnetizations in adjacent domains covered by a polymeric spacer layer. The liquid environment of the particle is a physiological buffer (indicated by red and blue ions +/-) to guarantee protein stability, while the colloidal stability is realized via specific surfactant concentrations (DS-).}%
\label{fig:1}%
\end{figure}
Although most transport concepts presented in past work have been performed in pure water,\cite{Ehresmann2015,Rampini2016,Yellen2007,Donolato2011} in realistic devices, proteins functionalizing the SPB surface  \textit{e.g.} as capture molecules would denaturate therein. This necessitates the use of physiological liquid environments, \textit{i.e.} buffer solutions, that preserve the correct protein structure which in turn directly determines the protein function. The physical characteristics of these physiological solutions are rather different from the ones of pure water, particularly their viscosities, permittivities, the ionic strengths and consequently the dependence of the surfaces' electrochemical potentials on the distance to the surface. This directly affects the surface forces which are mediated by the chosen dispersion medium. Hence, the dispersion medium does not only serve as a passive carrier liquid but plays an important role in the colloidal stability and for the surface sticking probability of the particles. As the delicate balance of surface forces, buoyancy, gravitation and magnetic forces determines the equilibrium separation between the beads and between bead and substrate, the described properties must be taken into account when aiming at bead transport without sticking and agglomeration. 
In past work, individual routines have been established to experimentally minimize agglomeration and adsorption of particles to the substrate surface: 
The distance between particle and magnetic layer has been increased by the introduction of a spacer layer\cite{Tierno2008,Yellen2007,Wirix-Speetjens2005,Donolato2011,Lim2017,Holzinger2015} which at the same time modifies the surface potential with the choice of material. This lowers the attractive magnetic forces and modifies the surface forces. Moreover, surfactants have been added in a specific concentration to reduce particle agglomeration and surface adhesion.\cite{Gunnarsson2005,Johansson2010,Ouk2017} In these approaches, the individually chosen parameters are usually restricted to the specific setup - consisting of the magnetic substrate, the SPB type, the immobilized proteins and the liquid - complicating a comparison of the studied transport concepts.

In order to understand the transport of biofunctionalized superparamagnetic beads (BSPBs) through physiologically relevant liquids more quantitatively, the current work describes a model to predict particle velocities for a prototype transport technology, where BSPBs are actuated by external magnetic field pulses superposed on periodic magnetic stray field field landscapes.\cite{Holzinger2015} This results in a transport of BSPBs in steps from domain wall to domain wall of the underlying topographically flat magnetic pattern (s. fig. \ref{fig:1}) with characteristic average bead velocities within the transport steps.    

This model includes a series of factors affecting the particle, from the distance dependent magnetic stray fields to the impact of surfactants on the viscous properties of the liquid and the surface forces determining the particle substrate distance. The model represents a basis to devise optimum conditions for the transport of BSPBs via the choice of surfactant and its concentration, the design of the magnetic stray field landscapes and the respective spacer thickness and substrate surface material. In figure \ref{fig:1}, the model system for this study is depicted, where the GFP mut2 structure (PDB code: 6GQG\cite{Lolli2018}) image was generated using the PyMOL Molecular Graphics System. The magnetic stray field landscape is fabricated according to ref. \citen{Holzinger2015} and as a spacer a 700~nm thick layer of bio-compatible poly(methyl methacrylate) (PMMA) was spin coated onto the magnetic thin film system. As prototype BSPBs we chose to immobilize the green fluorescent protein (GFP) on the surface of SPBs with a nominal diameter of 1 $\mu$m (see exp. methods section). 

GFP functionalized BSPBs are not only advantageous for particle tracking by the GFP fluorescence, but GFP itself is a sensitive probe for the detection of denaturation processes caused by the surrounding medium. The denaturation of GFP would lead to an unfolding of its beta-barrel structure\cite{Lolli2018} such that its spectral properties will change.\cite{Sokalingam2012} Moreover, its stability is comparable to proteins like antibodies, which are well-established high-affinity capture molecules for biomoarker detection.\cite{Pavoor2009} For the current experiments we varied the surface forces systematically by the concentration of the amphiphilic surfactant sodium dodecyl sulfate (SDS), of which data is well available,\cite{Khademi2017,Lee2011,Nielsen2007,Saeed2009,Sokalingam2012} in phosphate buffered saline (PBS). The average BSPB transport velocity (within the above mentioned steps of the stepwise transport) has then been determined as a function of SDS concentration and at the same time the protein's function has been monitored by the remaining fluorescence over time scales of days. Using the experiments in comparison to the model it is possible to determine parameter spaces for realistic lab-on-chips using magnetic particles.

%% file: Model.tex
\textbf{Magnetic field landscape} - 
The magnetic field landscape (MFL) over an exchange biased thin film patterned into magnetic parallel stripe domains with head-to-head and tail-to-tail domain configuration\cite{Ehresmann2004} (see fig. \ref{fig:1}) has been modeled using the object oriented micromagnetic framework (OOMMF) package\cite{oommf99}. The magnetic stray field landscape (MFL) with components $H_\mathrm{x}(x,y,z)$ and $H_\mathrm{z}(x,y,z)$ was afterwards computed as the superposition of the magnetic fields from all magnetic moments obtained from the simulation in dipole approximation.	Due to the stripes' parallel geometry, the transport is independent of the $y$ coordinate. The magnetic field landscape 
determined in this way (more details given in Suppl.) can similarly be determined for other kinds of patterned magnetic substrates.

\begin{figure}[!t]%
\includegraphics[width=\columnwidth]{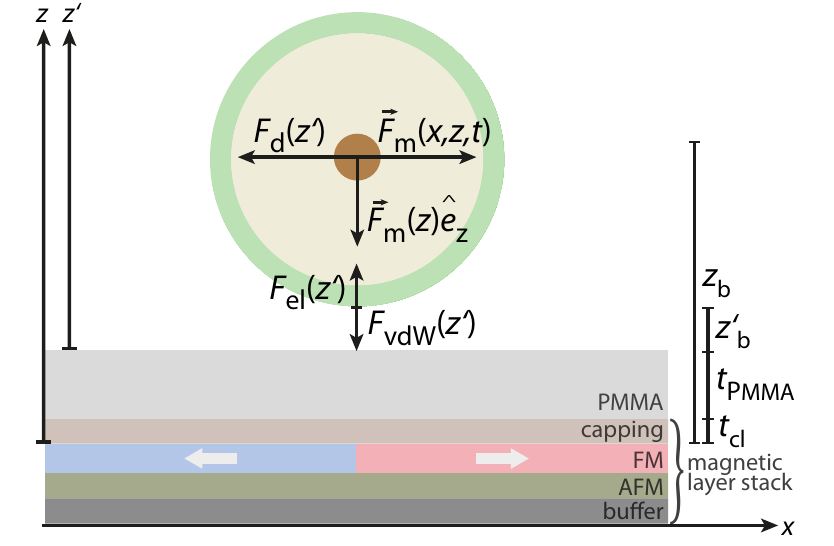}%
\caption{Sketch of the relevant forces on a BSPB (indicating the GFP surface functionalization in light green and the magnetic core in brown) for the theoretical description. Along the vertical axis, the electrostatic ($F_\mathrm{el}(z')$), van-der-Waals ($F_\mathrm{vdW}(z')$) and magnetic force ($\vec{F}_\mathrm{m}(z)\hat{e}_\mathrm{z}$) are indicated from which balance the steady-state distance is derived. The distance dependent forces determining the particle velocity are pointing horizontally: the drag force ($\vec{F}_\mathrm{d}(z')$) and the time dependent magnetic force ($\vec{F}_\mathrm{m}(x,z,t)$) which actuates the bead. The magnetic substrate consists of a buffer layer, an EB bilayer (antiferromagnet + ferromagnet) which is covered by a capping layer of the thickness $t_\mathrm{cl}$. Please note, that $z=z'+t_\mathrm{cl}+t_\mathrm{PMMA}$, with the spacer thickness $t_\mathrm{PMMA}$.}%
\label{fig:2}%
\end{figure}
\textbf{Steady-state distance} - 
The forces which are relevant for the theoretical description of magnetic bead transport in the model system are sketched in figure \ref{fig:2}. The two vertical axes in this depiction account for the fact that the individual forces are exerted either at the particle surface or at its center. The steady-state distance $z'_\mathrm{b}$ between the lowermost part of the BSPB and the PMMA surface, \textit{i.e.} the bottom of the microfluidic container, results from balancing the magnetostatic force $F_\mathrm{m}(z)$ and the surface forces (commonly referred to as DLVO forces, after the names of Derjaguin, Landau, Verwey, Overbeek\cite{DL1941,VO1948}), namely the electrostatic force $F_\mathrm{el}(z')$ and the van-der-Waals force $F_\mathrm{vdW}(z')$. Gravitation and buoyancy are neglected in the present calculations, because they are both in the range of $10^{-15}$~N, which is orders of magnitude smaller than the other named forces. The separation $z'_\mathrm{b}$ between the surfaces results in a distance between the BSPB's center and the ferromagnetic layer of $z_\mathrm{b}=z'_\mathrm{b}+t_\mathrm{cl}+t_\mathrm{PMMA}+r$ with $t_\mathrm{cl}$ being the thickness of the magnetic thin film system's capping layer, $t_\mathrm{PMMA}$ the thickness of the PMMA spacer, and $r$ being the particle's hydrodynamic radius. 
The magnetostatic force is calculated by the gradient of the magnetic potential energy $U_\mathrm{b}$:\cite{Holzinger2015}
\begin{equation}
\begin{split}
\vec{F}_\mathrm{m}(x,z)&=-\vec{\nabla}U_\mathrm{b} \\\
&=-\mu_0(\vec{m}_\mathrm{b}(x,z)\cdot\vec{\nabla})\cdot\vec{H}_\mathrm{eff}(x,z)
\end{split}
\label{eq:magForce}
\end{equation}
where $\vec{H}_\mathrm{eff}(x,z)$ is the effective magnetic field, which is the superposition of the homogenous applied field $\vec{H}_\mathrm{ext}$ and the magnetic stray field of the magnetically patterned substrate $\vec{H}_\mathrm{MFL}(x,z)$, $\mu_0$ is the vacuum permeability and  $\vec{m}_\mathrm{b}(x,z)$ the magnetic moment of the BSPB which was calculated in point dipole approximation by the Langevin function (see Suppl.).\cite{Yoon2004} For the steady-state distance between particle and substrate surface only the vertical component of the magnetic force is relevant, \textit{i.e.} $\vec{F}_\mathrm{m}(z)\cdot\widehat{e}_\mathrm{z}$. The required field gradient has been determined from the $z$-dependent magnetic stray field values above the stripe domain pattern (equation 1, Suppl.). The values determined above the domain wall centers (averaged along the long stripe axis) have then been fitted to the function $H_\mathrm{z,fit}(z)=a+b\cdot z^c$. The magnetic force was subsequently calculated for the effective magnetic field $\vec{H}_\mathrm{eff}(z)\cdot\widehat{e}_\mathrm{z}=H_\mathrm{z,fit}(z)+H_\mathrm{ext,z}$, where $H_\mathrm{ext,z}$ is the $z$-component of the externally applied magnetic field.

The distance dependent electrostatic force as a part of the surface forces is described by\cite{Wirix-Speetjens2005}
\begin{equation}
\begin{split}
F_\mathrm{el}(z')=&\frac{2 \pi \epsilon_\mathrm{r}\epsilon_\mathrm{0} \cdot r \cdot \kappa}{1-\exp(-2\kappa z')} \cdot[2\zeta_\mathrm{b}\zeta_\mathrm{s}\cdot \exp(-\kappa z') \\\
&\pm(\zeta_\mathrm{b}^2+\zeta_\mathrm{s}^2)\cdot \exp(-2\kappa z')]
\end{split}
\label{eq:elForce}
\end{equation}
with the permittivity $\epsilon_\mathrm{r}\epsilon_\mathrm{0}$ of the liquid medium and the zeta potentials $\zeta_\mathrm{b}$ and $\zeta_\mathrm{s}$ of the BSPB and the substrate surface, respectively. We use the zeta potentials as an approximation for surface potentials as the latter can not be experimentally obtained. Hence, we also use the hydrodynamic rather than the nominal radius. The force can be calculated either using the lower sign in equation \ref{eq:elForce} assuming a constant potential or using the upper sign, which corresponds to the constant surface charge density assumption. While it is known that neither of the two quantities remains constant during the approach of both surfaces\cite{Israelachvili2011}, the two approaches can be used as limiting cases. The electrostatic force furthermore depends on the Debye-H\"uckel inverse double-layer thickness 
\begin{equation}
\kappa=\sqrt{\frac{2000\cdot N_\mathrm{A}\cdot e^2\cdot I}{\epsilon_\mathrm{r}\epsilon_\mathrm{0}\cdot k_\mathrm{B}T}}
\label{eq:kappa}
\end{equation}
with the Avogadro constant $N_\mathrm{A}$, Boltzmann's constant $k_\mathrm{B}$, the temperature $T$ and the elementary charge $e$. As $\kappa$ depends on the ionic strength of the liquid $I=1/2\sum_{i}c_i\cdot z_i^2$, the strong influence of the concentration $c_i$ as well as the valency $z_i$ of each ion species $i$ in the given dispersion medium on the electrostatic force becomes obvious. For the present model, all ions of the buffer and the surfactant in the chosen concentrations need to be considered which yields $I=$172~mM for the PBS buffer without additional SDS. Calculating $F_\mathrm{el}(z')$ requires furthermore a value for the sphere's hydrodynamic radius ($r=$(614$\pm$22)~nm) and its zeta potential (see fig. \ref{fig:velcoities}), which were experimentally obtained for the utilized BSPBs in PBS via dynamic light scattering. The given uncertainties are single standard deviations from the mean value of the measurements. For each surfactant concentration, $\epsilon_{\mathrm{r}}$ of the dispersion medium PBS as well as $\zeta_\mathrm{s}$ for PMMA were estimated from computational as well as experimental results\cite{Khademi2017}. Note that this is an approximation since the values extracted from the cited article are based on a pH of 7 and an ionic strength of 100~mM, while we worked at pH$=$7.4 and at ionic strengths above 170~mM.
While the electrostatic force is repulsive for same zeta potential signs as it is the case in our experiment, the van-der Waals force $F_\mathrm{vdW}$ as the second DLVO force, is attractive.\cite{Gregory1981} Under consideration of the hydrodynamic radius and the Hamaker constants for the involved materials the distance dependence of $F_\mathrm{vdW}$ between a sphere and a plane was determined (see Suppl.).

After balancing the three forces $F_\mathrm{m}(z)$, $F_\mathrm{el}(z')$ and $F_\mathrm{vdW}(z')$ for a particle positioned above a domain wall, the derived steady-state distance $z'_\mathrm{b}$ with an uncertainty from Gaussian error propagation can be used to theoretically predict the particle's trajectory during a transport experiment. 

\textbf{Particle trajectory \& steady-state velocity} - 
The forces governing the transport velocity are the position ($x$,$z$) dependent magnetostatic force $F_\mathrm{m}$ causing the bead's actuation and the $z'$ dependent drag force $F_\mathrm{d}$ exerted by the liquid. For the magnetic actuation, equation \ref{eq:magForce} will then be evaluated for the x-component of the motion. As we use parallel stripe domains along the $y$ direction the potential is flat along $y$ such that $dU/dy = 0$. In order to correctly describe $H_\mathrm{eff}(x,z,t)$ and consequently $m_\mathrm{b}$, the magnetic field landscape is calculated in the determined distance $z_\mathrm{b}$ and the external field $H_\mathrm{ext}(x,z,t)$ is described in its temporal evolution during the experiment. The applied trapezoidal magnetic field pulses in $x$ and $z$ direction used in this study cause a dynamic transformation of the bead's potential energy landscape which drives their step wise transport.\cite{Holzinger2015}
The opposing drag force
\begin{equation}
F_\mathrm{d}(x,z')=6\cdot \pi \cdot r \cdot \eta_\mathrm{l}(c_\mathrm{SDS}) \cdot f_\mathrm{d}(z') \cdot \vec{v}_\mathrm{b}(x,z')
\label{eq:dragForce}
\end{equation}
is given by Stokes law for low Reynolds numbers, since they represent the utilized microfluidic device in good approximation. Here, $r$ is the bead radius, $f_\mathrm{d}$ is the distance dependent drag force coefficient and $\eta_\mathrm{l}$ is the liquid's viscosity, which in our case is surfactant dependent.\cite{Liu2007} The SDS concentration dependence of the viscosity $\eta_\mathrm{l}$ was determined by a linear fit from the experimental data of ref. \citen{Khademi2017}. In the following we only inspect the average steady-state velocity for one motion step, since the time interval for the bead's acceleration is with 1~$\mu$s much smaller than all relevant timescales for the presented experiments and can therefore be neglected.\cite{Holzinger2015}
By balancing $F_\mathrm{d}(x,z_\mathrm{b}')$ and $F_\mathrm{m}(x,z_\mathrm{b})$ at a given time $t$, the spatial dependence of the BSPB's momentary steady-state velocity 
\begin{equation}
\vec{v}_\mathrm{b}(x,z',t)=-\frac{\mu_0(\vec{m}_\mathrm{b}(x,z,t)\cdot\vec{\nabla})\cdot\vec{H}_\mathrm{eff}(x,z,t)}{6\cdot \pi \cdot r \cdot \eta_\mathrm{l} \cdot f_\mathrm{d}(z')}
\label{eq:velocity}
\end{equation}
can be calculated.\cite{Wirix-Speetjens2005,Holzinger2015} When applying this approach to transport concepts with acceleration phases in the temporal resolution of the experiment the equation of motion has to be solved rather than utilizing this force balancing approach.\cite{Urbaniak2018} Due to the position dependent energy landscape, $\vec{v}_\mathrm{b}(x,z',t)$ varies along the bead's path. 
Performing the calculation of $\vec{v}_\mathrm{b}(x,z',t)$ for each time interval $\Delta t = 10~\mu$s during the transport experiment we can construct the step like particle trajectory $x_\mathrm{b}(t_{i+1})=x_\mathrm{b}(t_{i})+\vec{v}_\mathrm{b}(x_{\mathrm{b},i},z,t_i)\cdot \Delta t +x_\mathrm{rw}$. The distance $x_\mathrm{rw}$ is added in order to account for the bead's Brownian motion (see. Suppl.). 

For a comparison of the simulated trajectories with positional data of the particles experimentally determined from microscopic videos via a tracking procedure\cite{Dingel2021}, mean step velocities for both data sets have been calculated in the same way. 
As the experimental data has been partly noisy, it was not possible to determine the particle velocities from the derivatives of the determined particle positions. 
Instead, the observed trajectories have been modeled by fitting a Gaussian error function to each step of the trajectory $x(t)$. The velocity was deduced from the function's derivative, which possesses by definition the shape of a normal distribution. The mean step velocity $\overline{v}_\mathrm{BSPB}$ was evaluated as the average velocity within the region around the maximum in which the values are above 50~$\%$ of the function's maximum (see. Supplementary). 
For the simulations, this procedure was also applied for the upper and lower boundaries of $z_\mathrm{b}$ derived from Gaussian error propagation in order to estimate the velocity's uncertainty for one step. This uncertainty together with the standard deviation from 5 averaged transport step simulations will be considered as an uncertainty in the evaluation (see fig. \ref{fig:velcoities}).

%% file: Results.tex
\begin{figure}[tb]%
\includegraphics[width=\columnwidth]{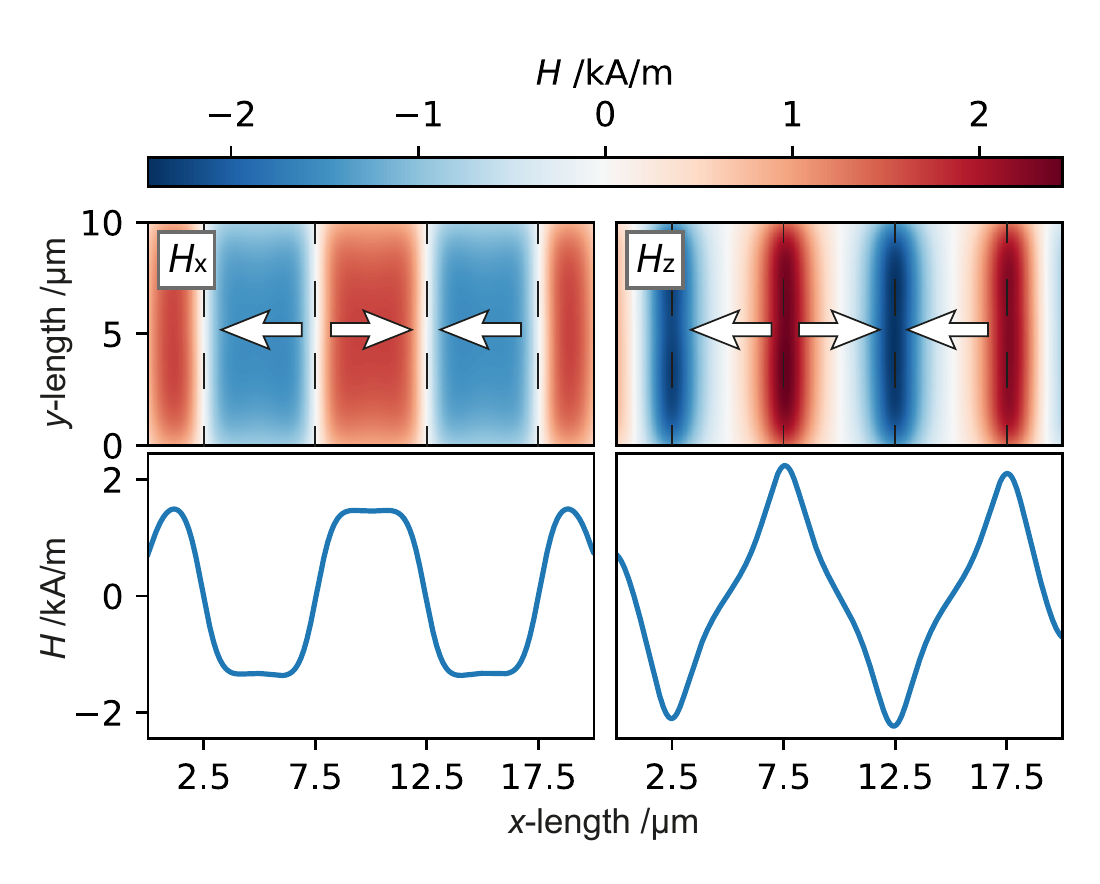}%
\caption{Top: The computed magnetic stray field components $H_\mathrm{x}$ and $H_\mathrm{z}$ in the $xy$-plane at $z_\mathrm{b}$=1.328 $\mu$m above the exchange biased stripe array. This distance corresponds to the steady-state distance $z_\mathrm{b}$ of the BSPB in PBS for $c_\mathrm{SDS}=1$~wt$\%$. The white arrows indicate the average remanent magnetization direction in the individual stripes separated by domain walls (dashed black). Bottom: respective line plot of the magnetic stray field components averaged along the $y$ coordinate.}%
\label{fig:mfl}%
\end{figure}
\textbf{Magnetic stray field landscape} - 
In figure \ref{fig:mfl} the $x$ and $z$ component of the magnetic stray field emerging from the parallel-stripe domains with in-plane head-to-head and tail-to-tail magnetization configurations is shown as a result of the OOMMF simulations. The field is shown for a distance above the ferromagnetic layer of $z_\mathrm{b}$=1.328~$\mu$m, which is the steady-state distance between the particle and the magnetic domain pattern at the lateral location of a domain wall. The inflection points in the magnetic stray field's $x$ component and the maxima and minima of the $z$ component coincide at the positions of the domain walls, therefore determining the potential energy minima of the magnetic beads. The stray field gradient along $z$ above the domain wall in superposition with the externally applied field  $H_\mathrm{ext,z}$ was subsequently used for the determination of the $z$ component of the magnetic force for the steady-state distance evaluation in combination with the interplay of the surface forces that both depend on the distance and the surfactant concentration in multiple ways.\newline
\begin{figure*}
\includegraphics[width=\textwidth]{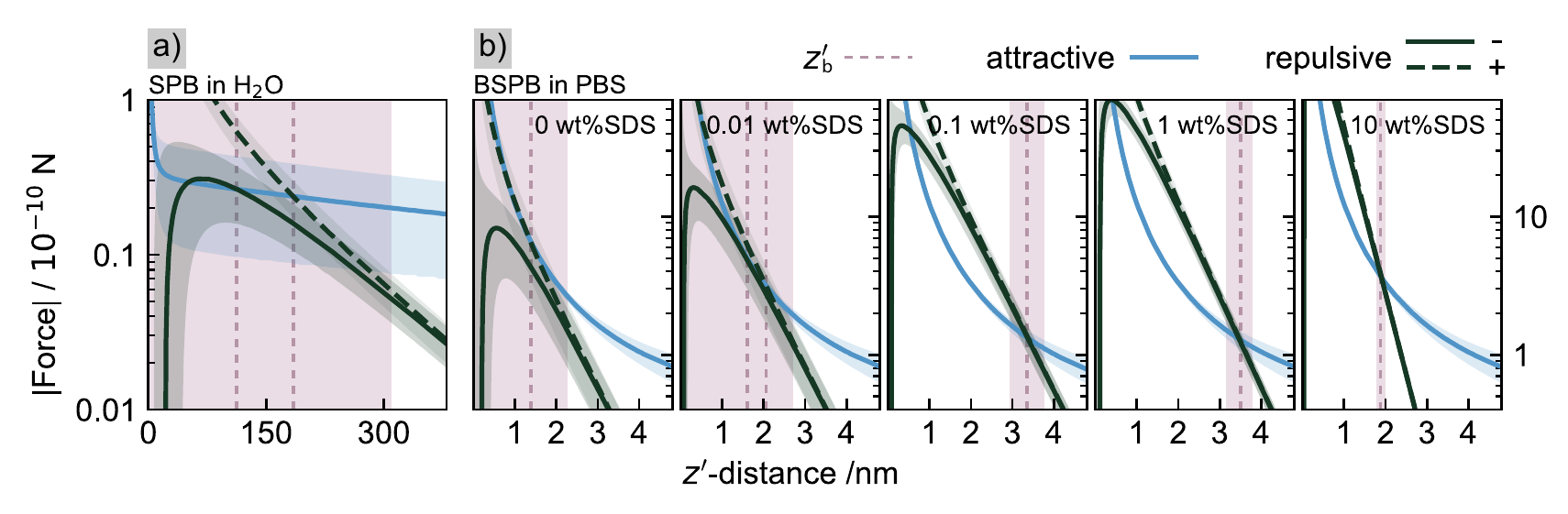}%
\caption{Absolute values of the repulsive and attractive forces between bead and substrate. The sum of the magnetic and the van-der-Waals forces (solid blue) is attractive. The two cases of either constant potential ($-$ in eq. \ref{eq:elForce}, solid dark green) or constant charge density ($+$, dashed dark green) for the electrostatic force are repulsive. The shaded areas in blue and dark green depict the corresponding uncertainties. While the left panel (a) depicts the force situation for a non-functionalized SPB in water, the graphs in the right panel (b) show the respective forces for BSPBs in PBS under the given SDS concentration. Note the different scales for the force and the distance. The vertical short-dashed lines indicate the steady-state distances for the two limiting cases of the calculated electrostatic force. The pale rose shaded area represents the uncertainty range for the equilibrium distance, derived from the intersections of the forces' error margins. When two dotted lines are given, the intersections do not coincide for the two limiting cases of the repulsive force, hence an averaged steady-state distance was determined.}%
\label{fig:forces}%
\end{figure*}
\textbf{Steady-state-distance \& transport velocities} - 
The interplay of forces will be discussed for BSPBs immersed in phosphate buffered saline (PBS) with an ionic strength of about 0.15~M with a variation of the surfactant concentration ($c_\mathrm{SDS}=[0,0.01,0.1,1,10]$~wt$\%$) in comparison to a plain SPB in water that was not previously functionalized with GFP. The sum of the magnetic force and the van-der-Waals force between particle and substrate surface is attractive (see fig. \ref{fig:2}), and it is displayed together with its uncertainty in figure \ref{fig:forces}.
On the other hand, the electrostatic force causes the repulsion of the particle. The two limiting cases of the calculated repulsive force are also shown in Fig.4. For all forces, the uncertainties were considered by a Gaussian error propagation. When the beads are immersed in the liquid during the experiment, they sediment first due to gravity and soon due to the additional attractive magnetic force until they reach a region (around $z'_\mathrm{b}$=6.4~nm for BSPB in PBS) in which the van-der Waals interaction dominates the attraction. Approaching now even smaller distances, the steady-state distance $z'_\mathrm{b}$ is reached when the repulsive force balances the attractive forces, which is indicated by the dashed vertical lines in figure \ref{fig:forces}. In the case that the two limiting cases for the calculated electrostatic force do not coincide, two dashed lines are given and the average value is taken as the effective steady-state distance. Similarly, the respective error margins of $z'_\mathrm{b}$ are derived from the crossing of the forces' uncertainties depicted as shaded regions, where the outer boundaries of both cases ($+/-$) were taken. While we retrieve an averaged steady-state distance of the SPB in water at $z'_\mathrm{b}=$148.5~nm within the error margins [0, 309]~nm, the separation between a BSPB and a PMMA surface in PBS has been calculated to be much smaller, \textit{i.e.} is at $z'_\mathrm{b}=$1.3~nm within the error margins of [0,2.3]~nm. This value is even smaller than the typical surface roughness of spin coated PMMA, which was experimentally determined to be in the range of 2~nm.\cite{Semaltianos2007} Under these conditions it is very likely that beads will stick to the PMMA surface.
Moreover, as the van der Waals force dominates for small distances, for the interaction among the beads, the van der Waals and electrostatic forces in PBS do not cancel each other for finite particle - particle distances. In this case the colloidal particle system is not stable, with BSPBs forming agglomerates with others in their vicinity. This corresponds to the experimental findings, where the BSPBs could not be actuated to perform full transport steps but were seen immobile or wiggling around a position, while some also formed agglomerates. 

\begin{figure}[!t]%
\includegraphics[width=\columnwidth]{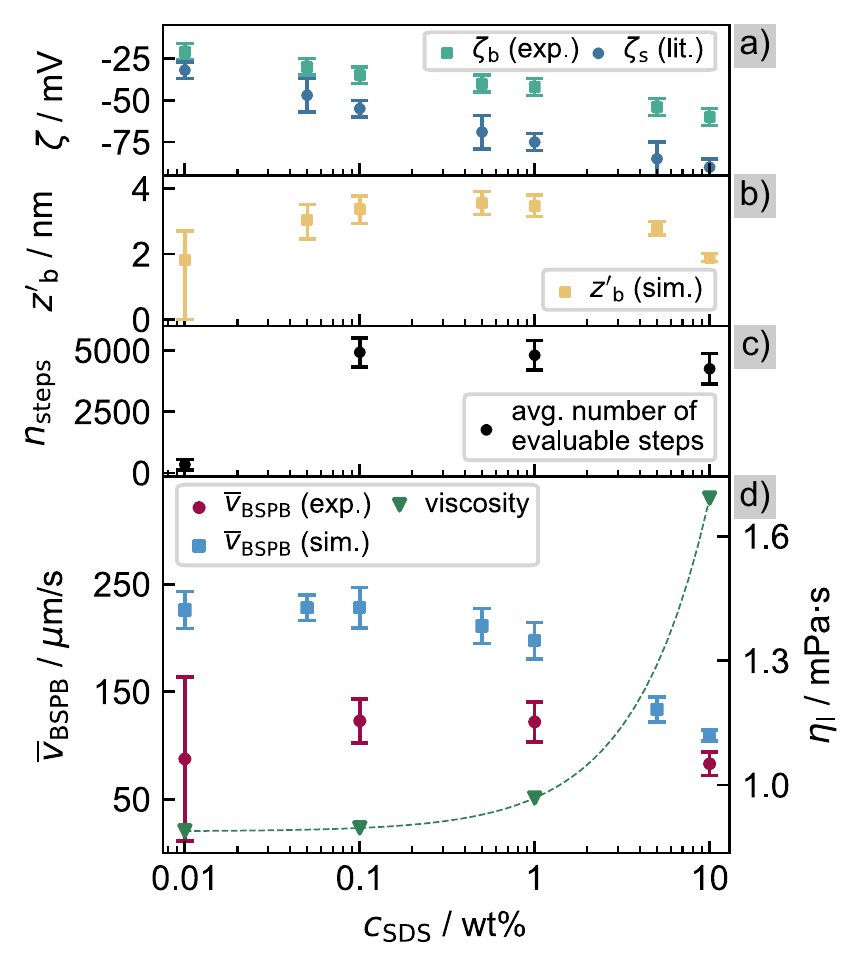}
\caption{Effects of the surfactant concentration on the transport of BSPBs. a) Zeta potential values $\zeta_\mathrm{s}$ of the PMMA substrate\cite{Khademi2017} and $\zeta_\mathrm{b}$ of the prepared BSPBs experimentally obtained by dynamic light scattering. b) Steady-state separations $z'_\mathrm{b}$ between the substrate surface and the BSPBs derived from the force evaluation (fig. \ref{fig:forces}). c) Average number of steps $n_\mathrm{steps}$ evaluable from the transport videos. d) Experimental results and theoretical values of the mean step velocities $\overline{v}_\mathrm{BSPB}$ during BSPB transport at room temperature.
The SDS concentration dependent progression of the viscosity $\eta_\mathrm{l}$ taken from ref. \cite{Khademi2017} is drawn with the respective axis on the right.}
\label{fig:velcoities}
\end{figure}

\begin{figure}[tb]%
\includegraphics[width=\columnwidth]{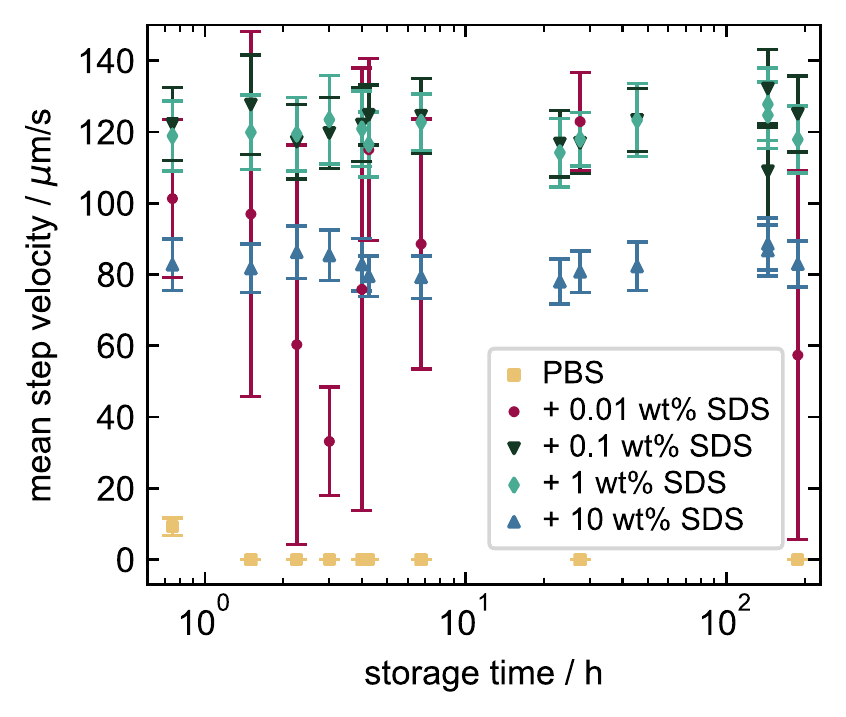}%
\caption{Experimentally obtained mean step velocities of GFP-functionalized BSPBs in buffer media with different concentrations of SDS at room temperature depending on the time after the BSPBs had been transferred to the respective storage buffer subsequent to the GFP functionalization. For each transport experiment, the fluorescence was simultaneously qualitatively detected indicating no unfolding of GFP for all storage times.}%
\label{fig:storage}%
\end{figure}

The interplay of the zeta potential, the permittivities and the ionic strength of the medium such as the used buffer, PBS, mimicking a physiological environment strongly influences the electrostatic force. 
Dynamic light scattering has been used to determine the zeta-potentials $\zeta_\mathrm{b}$ of the beads. They had been negative for all experimental cases (fig. \ref{fig:velcoities}a) due to the carboxylic functionalization of the SPBs and the fact that GFP has its isoelectric point (4.6-5.4)\cite{Chalfie2006} below the present pH value of 7.4. On the other hand, the absolute value $|\zeta_\mathrm{b}|$ of a BSPB in PBS is with 15~mV considerably smaller than $|\zeta_\mathrm{b}|$ of a SPB in water with 69~mV, which is due to electrochemical shielding by the ions in the buffer. Secondly, the Debye length $1/\kappa$ as a measure for the extension of the electrochemical double layer was determined to be 0.74~nm for PBS (see eq. \ref{eq:kappa}), while $1/\kappa=$100~nm is usually approximated for water as a medium due to ionic impurities.\cite{Butt2003} 
In order to avoid particle adhesion on the PMMA surface, SDS as an amphiphilic species was added in different concentrations, where the adsorption of the alkyl chain side of dodecyl sulfate anions (DS-) on the surfaces in the system causes a change of the electrochemical double layer associated with these surfaces, \textit{i.e.} increases their negative charge density. This results in increasing absolute zeta potential values for increased surfactant concentration, as can be seen in the experimental data as well as in literature (see fig. \ref{fig:velcoities}).\cite{Khademi2017} (A possible contribution of so-called non-DLVO forces to the steady-state distance consideration is assumed to be negligible in our system, since the beads' surface is hydrophilic (mostly negatively charged GFP on the BSPBs at our pH / COO- terminated SPBs) and the PMMA surface is partly hydrophobic, so that neither attractive hydrophobic nor repulsive hydration forces are expected to contribute.) Under the addition of different concentrations of SDS, a transport was experimentally achieved, which was in agreement with the SDS concentration dependent computed steady-state distances. 
The received values for $z'_\mathrm{b}$ extracted from the force curves in figure \ref{fig:forces} with the respective errors were subsequently used for the step velocity computation and are given in figure \ref{fig:velcoities} b). The increasing absolute zeta potentials cause an increase of the steady-state distance for SDS concentrations of up to 0.5~wt$\%$. However, $z'_\mathrm{b}$ drops beyond this concentration because the course of $F_\mathrm{el}$ becomes steeper with the increase of the Debye length $\kappa$ due to the solutions' ionic strength for higher $c_\mathrm{SDS}$. Correspondingly, the results for the mean step velocities $\overline{v}_\mathrm{BSPB}$ obtained via the model 
as well as in the concluded experiments 
are also presented as a function of the SDS concentration. Note, that the given velocities are averaged experimental values including acceleration and deceleration phases of evaluable transport steps, disregarding the number of beads sticking to the surface.
As a measure for the transport efficiency, we furthermore plot the averaged number of evaluable steps $n_\mathrm{steps}$ per SDS concentration in figure \ref{fig:velcoities}, while for each concentration 10 to 14 videos were recorded at different times after the biochemical coupling of GFP with a fixed particle concentration in the dispersion (see fig. \ref{fig:storage}).

Although the velocity of a BSPB at a SDS concentration of 0.01~ wt$\%$ with a steady-state distance of $z'_\mathrm{b}=1.8~$nm (error margins [0,2.7]~nm) was theoretically obtained to be around 223~$\mu$m/s it can be expected that this value is too high. Firstly, because the z-distance lower error margin is zero, which speaks for the particle being adhered. Secondly, because the distance of 1.8~nm is still smaller than the surface roughness of spin coated PMMA,\cite{Semaltianos2007} similar to the case of BSPBs in PBS without surfactant. Hence, the particles will possibly not be able to move in this close proximity as it was also seen in the experiments. Accordingly, the number of successfully evaluated steps as a measure for the quality of the transportation in this SDS concentration of 0.01~wt$\%$ is only 7~$\%$ of the steps evaluated for the concentration of 0.1~ wt$\%$. This low transport efficiency gives rise to a large experimental uncertainty of the step velocity $\overline{v}_\mathrm{BSPB}=(87\pm 76)~\mu$m/s. 
For increased SDS concentrations the transport efficiency increases resulting in reduced error margins.

An explanation for the deviation between model and experiment could be a systematic error by overestimating the magnetic force that arises from the approximations made during the simulation of the magnetic layer properties. Particularly, the step-like transition between the simulated domains with opposing exchange bias field was previously shown to yield an overestimated charge density in the domain wall and consequently a higher magnetic stray field strength.\cite{Mitin2018,Zingsem2017} This results in the calculation of too high step velocities, although the steady-state distance will be unaffected for BSPBs as it is in the van der Waals dominated distance regime of some nm. (A reason for the stray field overestimation is that the slope of the resist mask walls during the ion bombardment patterning together with the scattering of ions are expected to cause a gradual change of anisotropy rather than an abrupt transition.\cite{Zingsem2017})

Nevertheless, the step velocity trend is qualitatively captured by the model: When increasing $c_\mathrm{SDS}$ from 0.1 to 1~wt$\%$ the theoretical velocity drops from 226~$\mu$m/s to 195~$\mu$m/s, firstly due to the enhanced separation $z_\mathrm{b}$ causing a loss in MFL strength and consequently a reduced actuating magnetic force and secondly because of the drag force increase caused by the enhanced viscosity at this SDS concentration (see right axis of figure \ref{fig:velcoities}d). For $c_\mathrm{SDS}=$10~wt$\%$, the reduced particle-MFL separation again enhances the magnetic force, however the effect of the enhanced viscosity is even stronger which causes the theoretical step velocity to be reduced to 108~$\mu$m/s. From both the experimental and theoretical results, there seems to be an optimum SDS concentration between 0.1 and 1~wt$\%$. Accordingly, experiments with 0.1~wt$\%$ SDS gave the maximum for the average number of evaluable steps in the respective videos. This is valid for the model system presented in this study and will vary for other combinations of magnetic bead, buffer solution, surfactant as well as spacer thickness and on the magnetic domain pattern used for the actuation. \newline
\textbf{Protein Stability} - Since the critical concentration for micellation (CMC) of SDS in water is 8~mM = 0.23~wt$\%$ and is expected to be lower in our buffer,\cite{Orwick-Rydmark2016} we would assume to observe a reduction of the fluorescence intensity because global and cooperative unfolding occurs around the CMC and results in a rapid denaturation.\cite{Nielsen2007, Lee2011}
However, non-immobilized GFP was also previously shown functional above the CMC which underpins its high stability:\cite{Saeed2009,Sokalingam2012} While a relative fluorescence above 85~$\%$ at a SDS concentration of 0.5~wt$\%$ for pH values above 7.5 is reported at room temperature,\cite{Saeed2009} the relative fluorescence was lowered to $\sim$40~$\%$ at elevated temperatures of 50~$^\circ$C.\cite{Sokalingam2012} In order to investigate the surfactant's influence on the stability of the immobilized protein, the BSPBs were stored in the same media that we used for the transport experiments at room temperature and the fluorescence was qualitatively tested in a fluorescence microscope while the simultaneous transport indicated a remained colloidal stability. At all chosen points in the time frame of 8 days, a fluorescence of the beads was detected for each storage medium and simultaneously the transport properties were evaluated via the mean step velocities as described earlier. As depicted in figure \ref{fig:storage}, we observed no significant changes in the mean step velocity for $c_\mathrm{SDS} = [0.1,1,10]~$wt$\%$ indicating a preserved colloidal stability and it can be expected that the steady-state distance between BSPB and PMMA surface remained constant for all transport experiments. 
For particles in PBS at 0.01~wt$\%$ SDS it has not been possible to determine a trend due to the small number of evaluable transport steps causing larger uncertainties than for the other concentrations. These findings are in correspondence with the simulated small steady-state distance and the respective uncertainty for this buffer medium for which the transport properties may vary depending on slight variations of the experimental conditions, \textit{e.g.} environmental temperature or the position on the PMMA-covered micromagnetic array.
Furthermore, we account the remained GFP functionality at SDS concentrations of up to 10~wt$\%$, which is considerably higher than in the reports of non-immobilized GFP,\cite{Saeed2009,Sokalingam2012} to an entropic effect due to the immobilization which hinders the unfolding of the protein. Hence, fewer unfolded conformations are available for the bound protein, which enhances the stability.\cite{KnottsIV2008} 

%% file: Conclusion.tex
We have theoretically and experimentally studied the influence of different surfactant concentrations on the transport of GFP functionalized superparamagnetic beads in PBS mimicking physiological conditions, because experimentally no actuation in pure buffer was possible. Therefore, the particles' steady-state velocities and the number of evaluable transport steps were investigated as a measure for the transport behavior. Our bead actuation concept combines a topographically flat and magnetically structured substrate with external magnetic field pulses. The introduced theoretical model represents not only the magnetic and the drag force acting on a particle at any given distance but also addresses the influence of surface forces in order to accurately predict the equilibrium bead-substrate separation distance in dependence of the surfactant concentration. The steady-state distance is an indispensable key information when simulating a bead's trajectory and velocity. The experimentally received trend for particle velocities as well as the transport efficiency measured via the number of evaluable transport steps matches the predicted tendency where the mobility is enhanced within a certain range of the surfactant concentration, which is for the presented system 0.1 to 1~wt$\%$ sodium dodecyl sulfate (SDS). 
The presented model can be applied to other bead transport concepts when the governing forces can be determined. 
Moreover, we demonstrated the effect of SDS on the transport efficiency over the timescale of 8 days, in which transportation remains successful, while coincidentally the fluorescence was qualitatively observed to remain. This indicates a preserved protein stability over days which can possibly be explained entropically by the reduced number of unfolded states due to the immobilization and promotes GFP as a candidate for similar studies due to its stability. The possibility to forecast conditions where particle agglomeration and their adhesion to the substrate are avoided together with the boundary condition of preserved protein functionality renders the presented model as basis for future lab-on-a-chip explorations.

We conclude that optimizing the transportation mechanism for non-functionalized beads in water and expecting the same transportation yield subsequent to the protein immobilization is a misleading approach for the design of a biosensing lab-on-chip unit. Instead, it is inevitable that the optimization of the biofunctionalized beads' properties, the properties of the liquid medium and the adaption of the transport characteristics goes hand in hand.

%% file: Experimental.tex
\textbf{Preparation of BSPB} - 
His tagged green fluorescent protein (GFP) mut 2 \cite{Cormack1996} was overexpressed in \textit{E. coli} BL21 DE3 RIL cells after induction with 1 mM IPTG over night at RT using the expression vector pET15b. 
Cells were harvested (7,000 xg, 10 min) and stored at -20$^\circ C$ until further processed. Cells were homogenized in TRIS-buffer (50 mM Tris(hydroxymethyl)-aminomethanhydrochlorid pH 8.0, 300 mM NaCl) plus EDTA-free protease inhibitor (cOmpleteTM, Roche) and lysed by French press (Thermo IEC). Lysates were centrifuged (40,000 xg, 45 min). Supernatant was passed through 0.45 mm sterile PVDF Rotilabo1-syringe filters (Roth) and applied to metal affinity chromatography (Protino1 Ni-NTA column 1 mL, Macherey-Nagel) using an FPLC system (Äkta) at 4$^\circ C$. Unspecifically bound proteins were removed with a 20 mM imidazole step, followed by elution of the His-tagged protein with an imidazole gradient (20 mM – 500 mM) in TRIS-buffer. Elution fractions (70 mM – 350 mM imidazole) were pooled according to their absorbance at 280 nm followed by extensively dialysed against PBS-buffer (phosphate buffered saline: 140 mM NaCl, 10 mM NaH$_2$PO$_4$ pH 7.4, 2.7 mM KCl, 1.8 mM KH2PO4).
Protein concentration was determined using absorbance at 488 nm (lmax) and the concentration calculated using an estimated extinction coefficient of 55,000 g/mol. Protein purity was controlled by Coomassie-stained SDS-PAGE\cite{Dynabeads}. GFP was soluble and stable, aliquots were stored at -20$^\circ C$. Dynabeads (Dynabeads\textsuperscript{\textregistered} MyOne\texttrademark Carboxylic Acid, ThermoFisher Scientific) conjugation was carried out in principal according to the suppliers’ protocol\cite{Laemmli1970}. Briefly, 75~µL of beads (10~mg/mL) were washed two times with cold (0~°C) 0.01~M NaOH and once with 50~mM MES-buffer (25~mM 2-(N-morpholino)ethanesulfonic acid pH~6.4). Beads were separated after each step with a magnetic apparatus for 3 min. After discarding the supernatant, the mixture was  sonicated after inserting new solution for 20~s to maintain a homogenous solution. Beads were activated with 25~mM N-hydroxysuccinimide (NHS, Sigma Aldrich) and 25~mM N-ethyl-N'-(dimethylaminopropyl)-carbodiimide (EDC, Sigma Aldrich) in MES with slow continuous shaking at RT for 30~min. After discarding the supernatant, the beads were washed twice with MES-buffer. 
GFP was diluted with PBS-buffer to a total concentration of 1, 10, 30, 60, 100~µM 
and added to the activated beads. As a negative control, coupling in absence of His-GFP was carried out. The beads were incubated on a shaker with slow constant speed at RT for 30~min. The conjugated beads were washed twice with MES-buffer plus 0.2~wt$\%$ Tween. For quenching non-reacted activated groups, a solution of 50~mM ethanolamine was added and incubated at RT for 15 min, followed by two washing steps with PBS-buffer plus 0.01~wt$\%$ Tween. The fluorescence intensities of the BSPBs were measured in a CLARIOstar\textsuperscript{\textregistered} microplate reader. Bead suspensions were diluted 1:50 and transferred to black 384 well plates (Brand). As a reference 200~nM GFP in PBS-buffer pH~7.4 was used. Each measurement was carried out in triplicate applying the well scan method (settings: 15 x 15 scanmatrix, 2 mm width, 8 flashes per scan point). For excitation and emission determination optical filters (BMG LABTECH) were used (Ex. 482-16~nm, Em.: 530-40~nm, Dichroic: LP 504~nm). From the comparison of the chosen GFP concentrations we determined the maximum in the relative fluorescence intensity for an initial GFP concentration of 60~µM (s. Supplementary fig. S4), which was subsequently used for the fabrication of BSPBs for transport experiments. Fractions of the BSPB suspension were transferred to the storage and transport media, namely PBS solutions with the SDS concentrations 10, 1, 0.1, 0.01 and 0~wt$\%$, and stored in the same buffer at room temperature afterwards. \newline
\textbf{Preparation of micromagnetic array} - 
The magnetic substrate we used for the actuation is an exchange bias layer system with artificial parallel stripe magnetic domains with head-to-head and tail-to-tail magnetization orientations in adjacent domains (figure \ref{fig:1}). This layer stack with the composition Cu$^\mathrm{10 nm}$/Ir$_\mathrm{17}$Mn$_\mathrm{83}^\mathrm{30 nm}$/Co$_\mathrm{70}$Fe$_\mathrm{30}^\mathrm{10 nm}$/Ta$^\mathrm{10 nm}$ was fabricated by rf sputter deposition and field cooled as described by Holzinger \textit{et al.} \cite{Holzinger2015}. The artificial parallel stripe domains were fabricated by ion bombardment induced magnetic patterning (IBMP)\cite{Lengemann2012,Juraszek2002} using a home-built Penning ion source. For this procedure, a lithographically deposited 700~nm thick photoresist (AZ1505, Microchemicals) parallel stripe structure (5~$\mu$m wide with a periodicity of 10~$\mu$m) served as a shadow mask for the ion bombardment. 10~keV He$^+$-ion bombardment was performed with an ion fluency of $2.2\cdot10^{15}$ ions/cm$^2$ in an applied magnetic field of 64~kA/m antiparallel to the initial EB direction. The bombarded EB system showed a magnetization reversal superposed of two hysteresis loops corresponding to the adjacent artificial magnetic domains with the absolute values of the opposing exchange bias fields $|H_\mathrm{EB,L}| =$ 12$\pm$1~kA/m ($H_\mathrm{C,L} =$ 5$\pm$1~kA/m) and $|H_\mathrm{EB,R}| =$ 7$\pm$1~kA/m ($H_\mathrm{C,R} =$ 3$\pm$1~kA/m). After the removal of the photoresist, a 700~nm thick layer of PMMA was coated on top of the EB layer system serving as a spacer unit.\newline 
\textbf{Transport experiments} - A microfluidic chamber corresponding to a fluid container volume of $\sim$3~$\mu$l was fabricated from Parafilm M and a microscopy glass slide. For the experiments suspensions of BSPBs [130 $\mu$g/ml (1:75 from stock)] in PBS buffer with SDS concentrations $c_\mathrm{SDS}$ between 0 and 10 w\% were prepared. After loading the chamber with the respective particle suspension, it was placed between two solenoid pairs for the creation of trapezoidal magnetic field pulses perpendicular (z) and parallel (x) to the sample surface plane (see fig. \ref{fig:1}).\cite{Holzinger2015} The plateau magnetic fields were chosen to be $H_\mathrm{x,max} =$ 2.4~kA/m ($H_\mathrm{z,max} =$ 4.0~kA/m), while the rise and fall had an alteration rate of $v_\mathrm{H}=$3.2 $\cdot 10^6$~A/(m s). The magnetic field pulses for both directions were supplied at a frequency of $\omega_\mathrm{ext} =$ 1.25~Hz, while the oscillations of $H_\mathrm{z}$ advanced $H_\mathrm{x}$ by $\pi$/2. The particle trajectories during this transport performed at room temperature were recorded via an optical microscope setup (40x optical magnification) equipped with a high-speed camera operated at a frame rate of 1000~fps (Optronis~CR450~x2) and subsequently analyzed with the tracking software AdaPT developed for particle tracking.\cite{Dingel2021}\newline 
\textbf{BSPB properties} - The same suspensions were used to obtain the hydrodynamic radius and the zeta potential of the BSPBs in a triplicate via dynamic light scattering with the \textit{Zetasizer Nano ZS90} (Malvern) at room temperature subsequent to 1~min ultrasonication.

%% file: Supplementary.tex
\textbf{Magnetic field landscape calculation} - 
The patterned exchange biased thin film with head to head and tail to tail domain configuration\cite{Ehresmann2004} was modeled using the object oriented micromagnetic framework (OOMMF) package\cite{oommf99}. Since the pattern has been fabricated by keV He ion bombardment, the magnetic properties differ in the adjacent domain types.\cite{Zingsem2017} Thus, for the saturation magnetization of the unbombarded domains an experimentally determined value is used while this value needs to be decreased for the bombarded stripes.\cite{Huckfeldt2017} Furthermore, the ferromagnet's uniaxial anisotropy is reduced accordingly and the relative changes were taken into account.\cite{Muglich2018} The exchange bias as the responsible anisotropy for the pinned magnetization direction of each stripe domain was accounted for in the OOMMF simulations by a fixed Zeeman term with exchange bias fields determined by magneto-optical Kerr measurements for each domain type. The sign of the Zeeman term has been inverted for the different stripe domain regions in order to appropriately implement the chosen head-to-head and tail-to-tail domain configuration. From the resulting magnetization values for the mesh elements of volume $V_i=(5~\mathrm{nm})^3$ in the chosen grid ($x= 20~\mu$m, $y= 10 \mu$m) the magnetic moment for each element $i$ was calculated as $m_\mathrm{x}=M_\mathrm{x}\cdot V_{i}$. The magnetic stray field landscape (MFL) at the position  $\vec{r}=\left(x,y,z\right)$ with components $H_\mathrm{x}(\vec{r})$ and $H_\mathrm{z}(\vec{r})$ was afterwards computed as the superposition of the magnetic fields from all magnetic moments obtained from the simulation in dipole approximation:
\begin{equation}
 \vec{H}(\vec{r})=\frac{1}{4\pi}\sum_{i}{\ \frac{3(\vec{R}\cdot{\vec{m}}_i)\ \vec{R}}{|\vec{R}|^5}}-\frac{{\vec{m}}_i\ }{|\vec{R}|^3}
\label{eq:dipolarfield}
\end{equation}
where $\vec{R}=\vec{r}-\vec{r_\mathrm{i}}$ is the distance vector between the observer position $\vec{r}$ and the position of the dipole (or here the mesh element) $\vec{r_\mathrm{i}}$.\cite{Nolting2013}

\textbf{Modeling a Trajectory Including Browninan Motion} - 
We described this random walk under the assumption of an infinite liquid environment as
\begin{equation}
x_\mathrm{rw} = \sqrt{\Delta t \frac{k_\mathrm{B}T}{3\pi\eta_\mathrm{l}f_\mathrm{d}(z)}}\cdot a_\mathrm{rand}
\label{eq:randomwalk}
\end{equation}
where $a_\mathrm{rand}$ is a random value sampled from a Gaussian distribution with a mean of 0 and a variance of 1. 

\textbf{The magnetic moment of a SPB} - 
The latter can be described in point particle approximation by the Langevin function \cite{Yoon2004}
\begin{equation}
\begin{split}
\vec{m}_\mathrm{b}(x,z)=m_\mathrm{s} \cdot \Biggl[&\coth(b\cdot \vec{H}_\mathrm{eff}(x,z))\\\
-&\left(\frac{1}{b\cdot \vec{H}_\mathrm{eff}(x,z)}\right)\Biggr]
\end{split}
\label{eq:magMom}
\end{equation}
where for the average saturation magnetic moment $m_\mathrm{s}$ a value of 2.45~$\cdot 10^{-14}$~Am$^2$ calculated from the vendor's information\cite{Monticelli2016,Dynabeads} was used and the Langevin parameter $b=$1.05$\cdot 10^{-4}$~m/A was taken from experimental data provided in ref. \citen{Lipfert2009}.

\textbf{Van der Waals Forces} - 
The distance dependence of $F_\mathrm{vdW}$ between a sphere and a plane writes as\cite{Gregory1981}
\begin{equation}
F_\mathrm{vdW}(z')=- \frac{A_\mathrm{H,123}\cdot r}{6z'^2}\left(\frac{1}{1+14z'/l_\mathrm{ret}}\right)
\label{eq:vdwForce}
\end{equation}
with the hydrodynamic radius $r$ of the sphere, the Hamaker constant $A_\mathrm{H,123}$ describing the interaction of the bead of material 1 and a substrate of material 2 in a liquid medium 3 via the relative permittivities $\epsilon_{\mathrm{r},i}$ and the optical refractive indices $n_i$ of the respective material/medium $i$. For the calculations $A_\mathrm{H,123}$ according to refs. \citen{Israelachvili2011} and \citen{Feldman1998} with the respective values for $n_i$ and $\epsilon_{\mathrm{r},i}$ of the materials 1 and 2, namely PMMA and polystyrene as an estimate for the BSPBs.\cite{Feldman1998}. The associated uncertainty is expected to be insignificant since the range of $A_\mathrm{H,123}$ from 1.2 to 1.8 $\cdot 10^{-20}$~J for the different surfactant concentrations used in the present experiments are in the range of Hamaker constants of proteins reported in literature (0.8 to 2.0$\cdot 10^{-20}$~J).\cite{Roth1996,Visser1972}
For the dispersion medium PBS the same SDS dependent values for $\epsilon_{\mathrm{r},3}$ as for the calculation of $F_\mathrm{el}$ were taken.\cite{Khademi2017} When approaching distances $z'_\mathrm{b}$ smaller than the radius of the sphere - like in this work-, retardation effects, which are reflected by the term in brackets, can be neglected.

\textbf{Evaluation of experimental data} - 
From the recorded videos we evaluated the beads’ trajectory via a tracking procedure. For this purpose, we used the particle tracking software AdaPT [Ref], which fulfills tracking in two subsequent steps: Locating the particles in the respective frames with subpixel accuracy and linking the extracted x-y coordinates into probable trajectories. The result of the tracking procedure is displayed in Fig. S1.
\begin{figure*}[!h]
\includegraphics[width=0.8\textwidth]{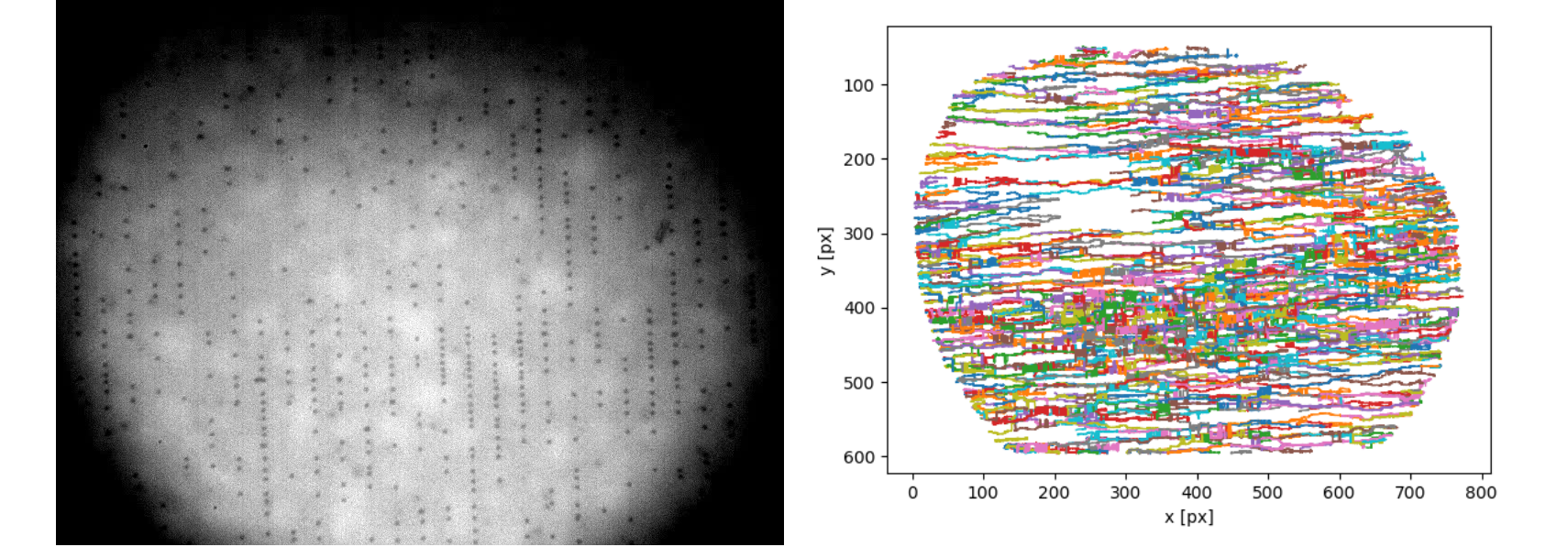}%
\caption{Left: exemplary video frame. Right: Output figure from tracking procedure. All trajectories of a video (8.1 s) in which the magnetic field sequence theoretically induced 20 steps for each particle.}%
\label{fig:video_frame}%
\end{figure*}

\newpage
\textbf{Steady-state velcoities} - As the tracked positional information of the beads from the videos can be noisy due to lighting conditions and the spatial resolution of the microscopic unit, we decided not to determine the momentary velocity for each frame from the numerical derivative. Instead, a Gaussian error function was fitted to each step within the observed trajectory x(t), where the velocity was deduced from the function’s derivative, which possesses by definition the shape of a normal distribution. In order to compare the experimental and the theoretical findings, we proceeded with both types of trajectory in the same manner.
Next, the mean step velocity was evaluated as the average velocity within the region around the maximum in which the values are above 50~\% of the function’s maximum. From the trajectories simulated via balancing magnetic and drag force during transportation time for different SDS concentrations, we deduced this method from the comparison with the momentary velocity and its average value for $v >$ 60~$\mu$m/s.
\begin{figure}[!h]
\centering
\includegraphics[width=0.8\textwidth]{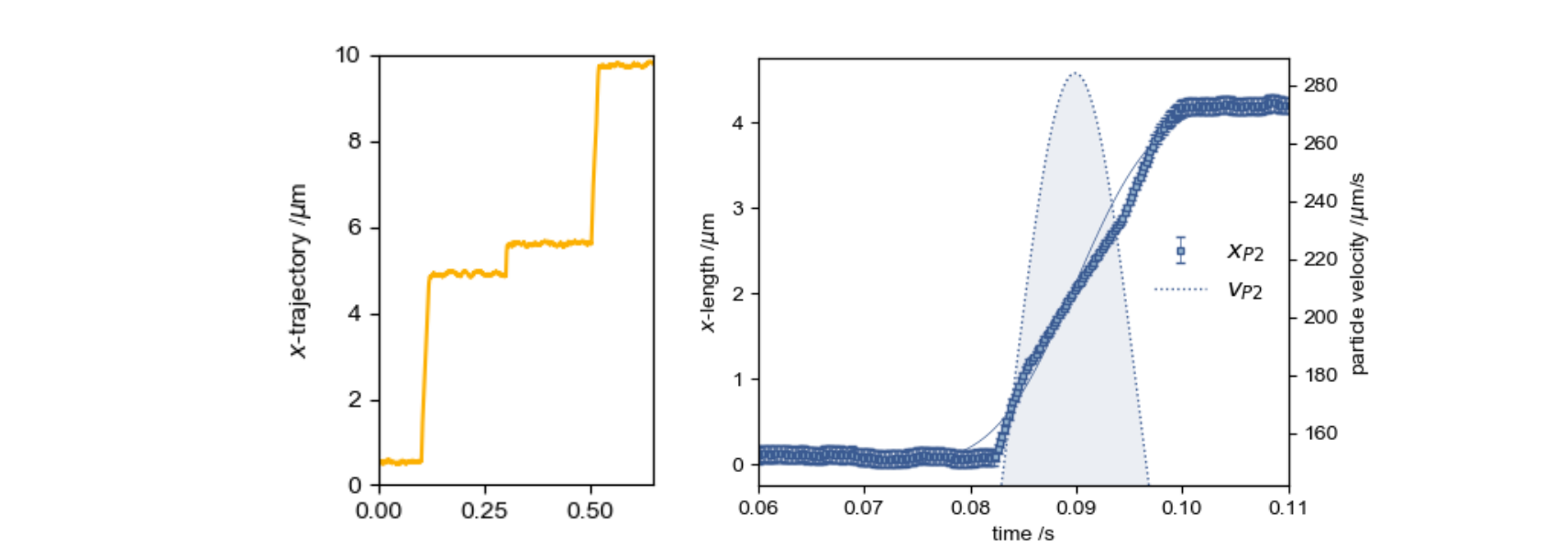}%
\caption{Evaluation of mean step velocity via error function fit to the large step within the simulated trajectory.}%
\vspace*{\floatsep}
\includegraphics[width=0.8\textwidth]{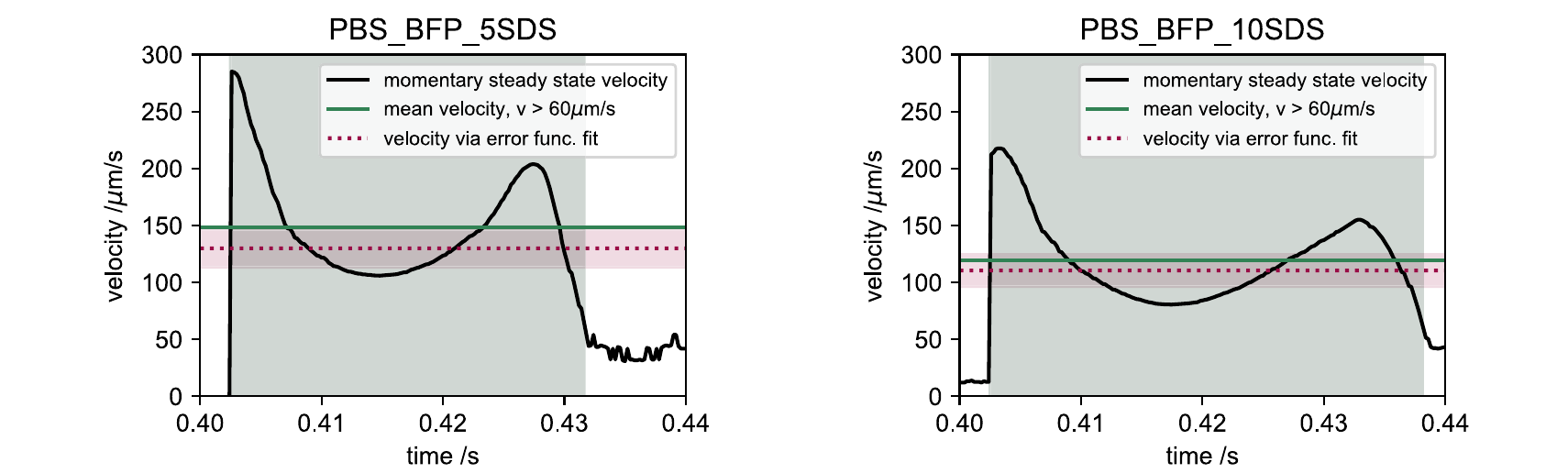}%
\caption{Simulated momentary steady-state velocities. Comparison between the momentary velocity, its mean value above a threshold of 60 µm/s (grey shaded region) and the result from the error function fit evaluation with the respective error.}%
\label{fig:method_velocity}%
\end{figure}
\newpage\
\textbf{Coupling efficiency} - Prior to the transportation experiments the coupling protocol for the biofunctionalization of SPBs with GFP was tested. While varying the concentration of GFP added to the suspension of beads and coupling reagents, we determined relative fluorescence intensities. Consequently, we chose a GFP concentration for coupling of 60~µM for all subsequent biofunctionalizations for the transport experiments.

\begin{figure*}[!h]
\includegraphics[width=0.9\textwidth]{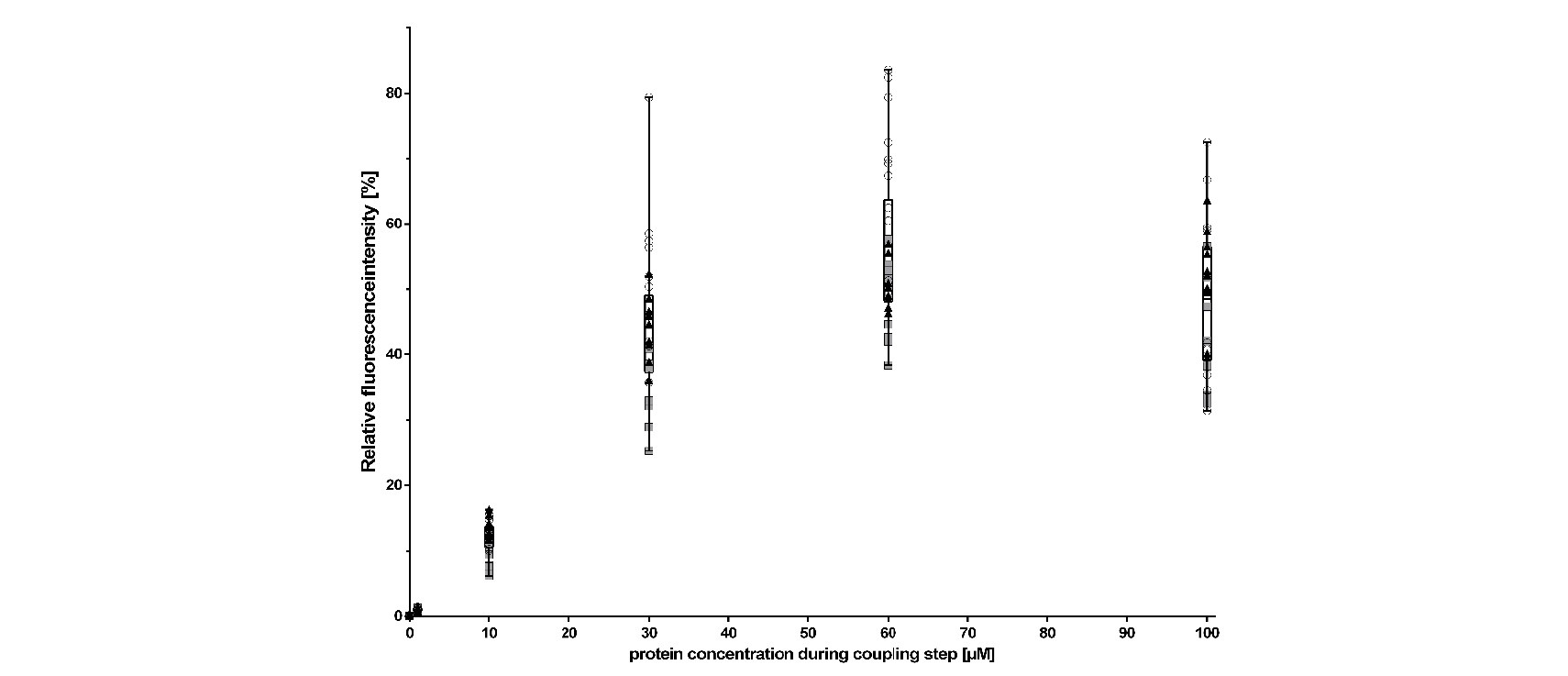}%
\caption{Determination of coupling efficiency in dependency of the protein concentration during the coupling process. Different concentrations of GFP, ranging from 1~µM up to 100~µM, were incubated with activated beads. After coupling, the GFP-fluorescence was measured with 100~\% set to the fluorescence of a 200~nM GFP solution. Measurements from three independent couplings (white circles, black triangles and grey squares) with 10 points were measured for each coupling.}%
\label{fig:GFP_coupling}%
\end{figure*}